\documentclass[fleqn,usenatbib]{mnras}
\usepackage{mathptmx}
\usepackage[varvw]{newtxmath}  
\usepackage[T1]{fontenc}
\usepackage{ae,aecompl}
\usepackage{times}
\usepackage{amsmath}
\usepackage{upgreek}
\usepackage{xcolor}
\usepackage{graphicx}
\usepackage{pdflscape}
\usepackage{multicol} 
\usepackage{amsmath,bm}
\usepackage{caption}

\begin{document}

\title[]{Connecting Star Formation in the Milky Way and Nearby Galaxies -II. An Observationally Driven Analytical Model for Predicting Cloud-Scale Star Formation Rates}

\author[J. W. Zhou]{
J. W. Zhou \thanks{E-mail: jwzhou@mpifr-bonn.mpg.de}$^{1}$
Amelie Saintonge $^{1}$
Sami Dib $^{2}$
Pavel Kroupa $^{3,4}$
\\
$^{1}$Max-Planck-Institut f\"{u}r Radioastronomie, Auf dem H\"{u}gel 69, 53121 Bonn, Germany\\
$^{2}$Max-Planck-Institut f\"{u}r Astronomie, K\"{o}nigstuhl 17, 69117 Heidelberg, Germany\\
$^{3}$
Helmholtz-Institut f{\"u}r Strahlen- und Kernphysik (HISKP), Universität Bonn, Nussallee 14–16, 53115 Bonn, Germany \\
$^{4}$
Charles University in Prague, Faculty of Mathematics and Physics, Astronomical Institute, V Hole{\v s}ovi{\v c}k{\'a}ch 2, CZ-180 00 Praha 8, Czech Republic
\thanks{E-mail: dib@mpia.de}
}

\date{Accepted XXX. Received YYY; in original form ZZZ}
\pubyear{2025}
\maketitle

\begin{abstract}
We construct a model by integrating observational constraints from the Milky Way and nearby galaxies to predict cloud-scale star formation rates (SFRs). In the model, we first estimate the initial total mass of clumps in a cloud based on the cloud mass, and then generate the initial clump population of the cloud using the initial clump mass function. Next, we model the star formation histories (SFHs) of the cloud to assign an age to each clump. We then sort out the intermediate-age clumps and calculate the total embedded cluster mass. Finally, we predict the SFR based on the duration of the embedded phase. The model-predicted SFR is broadly consistent with the observed SFR, supporting the plausibility of the model. The model primarily provides a theoretical framework that integrates a wide range of observational results, thereby clarifying the tasks for future observations.
\end{abstract}


\maketitle

\section{Introduction}\label{sec:intro}


Galaxies act as stellar nurseries, forming stars through the gravitational collapse of dense regions within molecular clouds. These clouds, which are widespread throughout galaxies, contain multiple clumps that serve as the local sites of star formation and the precursors of embedded star clusters
\citep{Kennicutt2012-50,Miville2017-834,Rosolowsky2021-502,Urquhart2022-510,Yan2017-607,Zhou2024PASP-1,Zhou2024PASP-2}.
The work of
\citet{Motte2018-56, Vazquez2019-490,Kumar2020-642, Henshaw2020-4, Zhou2022-514,Zhou2023-676,Zhou2024-686-146, Zhou2024PASA, Zhou2024-534,Zhou2025-537b,Zhou2025-699} (and reference therein) present a comprehensive multi-scale investigation of hub-filament structures, examining their morphology, kinematics, and evolution from dense core ($\sim$1000 AU) to clump ($\sim$1 pc), molecular cloud ($\sim$10-100 pc), and galaxy-cloud ($\sim$1000 pc) scales. These observations reveal hierarchical hub-filament structures with self-similar properties spanning sub-parsec to kiloparsec scales, highlighting their crucial role in the star formation process. Within this hierarchy, a dense core serves as a hub within a clump, a clump functions as a hub within a molecular cloud, and a molecular cloud acts as a hub within a galaxy. Hub-filament structures form through the gravitational contraction of gas structures. Velocity gradient analyses suggest that gas inflows along filaments are driven by gravity. Molecular gas is organized into network structures formed by the gravitational coupling of multi-scale hub-filament structures, where local gravitational centers (hubs) serve as the primary sites of star formation.

Progress in understanding the physical properties of giant molecular clouds (GMCs) is critical for elucidating the connection between interstellar gas dynamics and star formation on galactic scales.
Understanding what regulates the star formation rate (SFR) in molecular
clouds is a central problem in star formation theory.
Despite the short gravitational free-fall times of dense gas,
observations show that star formation is remarkably inefficient, with only a few percent of the gas converted into stars \citep{Evans2009-181,Kennicutt2012-50,Kim2022-516,Chevance2023-534,Zhou2025-541}.
A wide variety of theoretical models have been proposed to explain
this inefficiency, invoking turbulence, magnetic fields, global
gravitational collapse, and stellar feedback.
In turbulence-regulated models, supersonic (magneto-)hydrodynamic
turbulence generates a broad density distribution, and only gas in the
high-density tail becomes gravitationally unstable \citep{Krumholz2005-630,Padoan2011-730,Hennebelle2011-743,Federrath2012-761}.
Alternative models emphasize the time-dependent nature of molecular
clouds, proposing that star formation proceeds via global, hierarchical
collapse and accelerates as clouds evolve \citep{Zamora2014-793,Vazquez2019-490}.
Observationally motivated approaches link the star formation rate
directly to the mass of dense gas ($M_{\rm dense}$) above a column density threshold,
implying $\mathrm{SFR} \propto M_{\rm dense}$ \citep{Lada2010-724,Shimajiri2017-604}.
Feedback-regulated models argue that stellar feedback injects
momentum and energy that balance gravity on cloud scales, self-adjusting
the star formation rate, particularly in massive or high-surface-density
clouds \citep{Ostriker2011-731,Kruijssen2019-569}.

Recent advances in high-resolution, multi-wavelength observations now allow star formation to be resolved on the scale of individual molecular clouds in nearby galaxies.
In particular, high-resolution CO imaging from ALMA (Atacama Large Millimeter/submillimeter Array), along with other submillimeter observations, enables systematic investigations of the molecular cloud population beyond the Milky Way
\citep{Leroy2021-257,Leroy2021-255,Lee2022-258,Emsellem2022-659,Lee2023-944,Grishunin2024-682,Schinnerer2024-62}.
Since molecular clouds in the Milky Way can be resolved down to their internal structures, they provide a crucial reference for understanding the internal composition and star formation processes of molecular clouds in nearby galaxies. 
Star formation occurs primarily within clumps. 
Star formation observed on molecular cloud scales or larger is essentially an integrated outcome of the activity occurring within these clumps. 
To truly understand the underlying physics of star formation at those larger scales, it is essential to characterize the star-forming states of individual clumps. 
This requires turning to the Milky Way as a reference, leveraging surveys such as ATLASGAL and Hi-GAL \citep{Schuller2009-504,Urquhart2022-510,Molinari2010,Elia2021-504}, which are capable of resolving individual clumps, along with numerous ALMA follow-up studies that zoom in on single clumps in detail \citep{Sanhueza2019-886,Liu2020,Motte2022-662,Molinari2025-696}. 
In \citet{bridge}, we used CO (2–1) and CO (1–0) data cubes to identify molecular clouds and study their kinematics and dynamics in three nearby galaxies and the inner Milky Way. Molecular clouds in the same mass range across these galaxies show broadly comparable physical properties and similar star formation rates. Strong correlations were found between cloud mass and total clump mass, clump number, and the mass of the most massive clump \citep[see also][]{Zetterlund2019-881}. 

Unlike previous theoretical models, 
this work adopts a predominantly observationally driven approach. Rather than relying on specific theoretical assumptions, we attempt to synthesize existing observational results and treat the molecular-cloud-scale star formation rate as a natural consequence of a set of empirical relations. 
Clumps are identified as the true sites of star formation within molecular clouds. 
The star formation rate on molecular cloud scales, as a large-scale phenomenon, is in fact an integrated manifestation of star formation processes occurring on clump scales within molecular clouds.
As self-gravitating structures \citep{Liu2016-829, Urquhart2018-473, Evans2021-920}, clumps are almost decoupled from their surrounding large-scale environments \citep{Watkins2019-628A,Peretto2023-525,Zhou2023-676,Zhou2024-682-128} and therefore form stars in a nearly independent manner. Consequently, clumps, rather than molecular clouds, are taken to be the fundamental units of star formation. 
The structure of this paper is as follows:
In Section \ref{model}, we compile empirical relations at the molecular cloud and clump scales derived from observations of the Milky Way and nearby galaxies, construct the initial clump mass function of molecular clouds, and model their star formation histories. Section \ref{result} describes in detail the procedure for implementing the model predictions, presents an uncertainty analysis of the model parameters, and discusses the limitations of the model. Section \ref{conclusion} provides a brief summary.


\section{Model}\label{model}

\begin{table*}
\centering
\caption{Notation and symbol definitions used in the model.}
\label{symbol_table}
\begin{tabular}{ll}
\hline
\textbf{Symbol} & \textbf{Description} \\
\hline
$M_{\rm cloud}$ & Total mass of a molecular cloud \\
$M_{\rm clump}$ & Mass of an individual clump within a cloud \\
$M_{\rm clump,tot}$ & Initial total mass of clumps in a cloud (including past and present clumps) \\
$M_{\rm clump,tot,obs}$ & Observed total mass of clumps in a cloud (currently existing clumps only) \\
$M_{\rm clump,tot,m}$ & Total mass of clumps in the transition model between low- and high-mass clouds (Equation \ref{Mctio}) \\
$M_{\rm ecl}$ & Mass of an embedded star cluster \\
$\mathrm{SFR}_{\rm cloud,obs}$ & Observed star formation rate of a molecular cloud \\
$\mathrm{SFR}_{\rm cloud,p}$ & Model-predicted star formation rate of a cloud \\
$\Sigma_{\rm SFR}$ & Star formation rate surface density (SFR per unit area) \\
$I_{\rm 24 \mu m}$ & Total 24 $\mu$m flux within a cloud \\
$A$ & Projected area of a molecular cloud \\
$\mathrm{SFE}_{\rm clump}$ & Star formation efficiency of a clump \\
$\mathrm{SFE}_{\rm cloud}$ & Star formation efficiency of a cloud (stellar mass / cloud mass) \\
$\xi_{\rm clump}(M)$ & Clump mass function (CLMF) \\
$\beta$ & Power-law index of the clump mass function \\
$M_{\rm min}$ & Minimum clump mass \\
$M_{\rm max}$ & Maximum clump mass \\
$t_{\rm b}$ & Age of a clump \\
$t_{\rm b,p}$ & Peak of the clump age distribution (Gaussian SFH) \\
$\sigma_{\rm t}$ & Standard deviation of clump age distribution (Gaussian SFH) \\
$t_{\rm GMC}$ & Typical lifetime of a giant molecular cloud (GMC) \\
$t_{\rm emb}$ & Duration of the embedded phase of a cluster \\
$\overline{\rm CV}_i$ & Average coefficient of variation for parameter $i$ (measure of relative uncertainty) \\
$\overline{\Delta}_i$ & Average relative deviation of parameter $i$ from the reference SFR \\
$f_i$ & Fractional contribution of parameter $i$ to total uncertainty \\
$k_{\rm clump}$ & Normalization constant of the CLMF \\
\hline
\end{tabular}
\end{table*}

\subsection{Correlations from the Milky Way}\label{s1}

\begin{figure}
\centering
\includegraphics[width=0.475\textwidth]{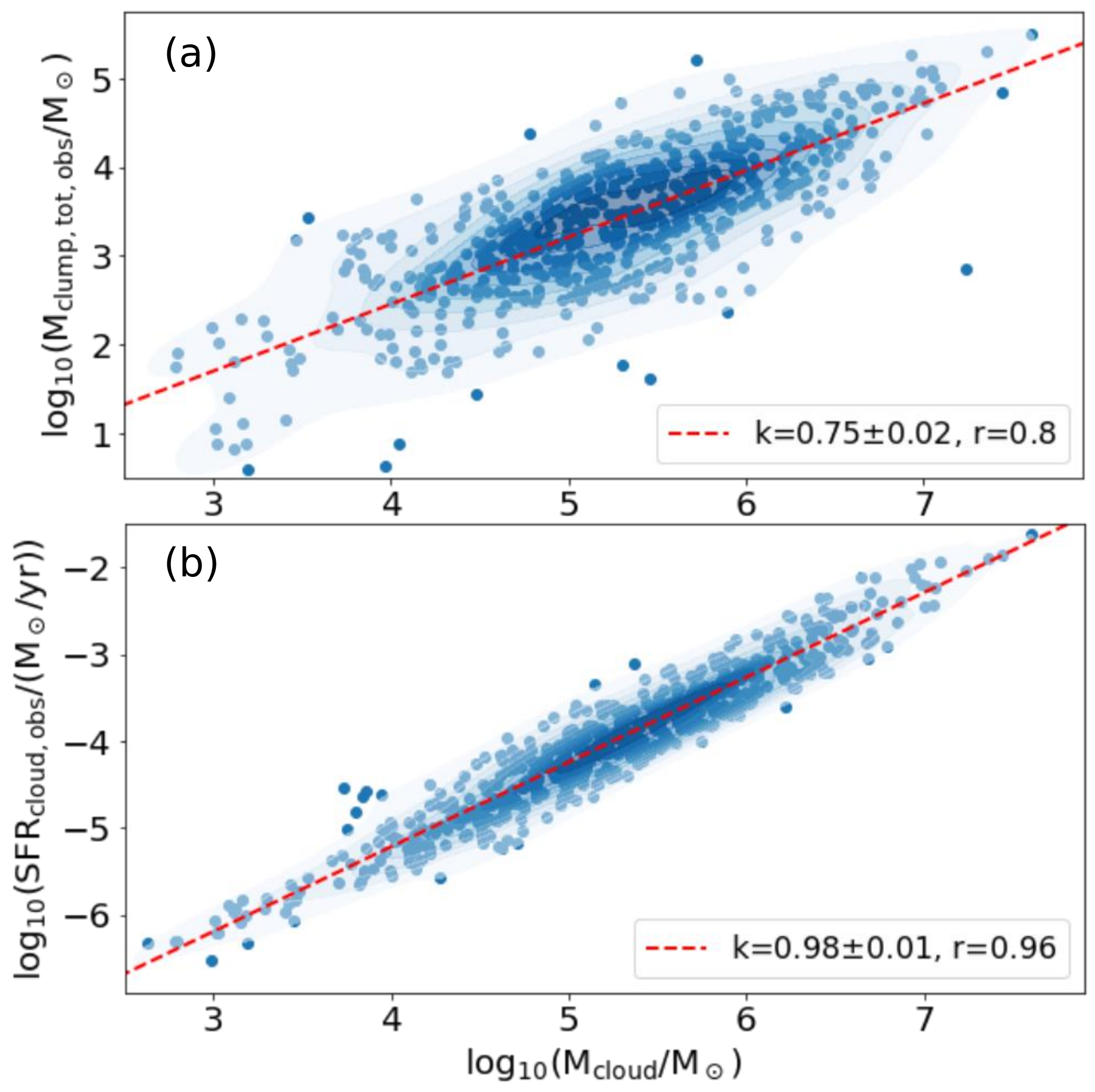}
\caption{
Correlations between the total clump mass in a cloud ($M_{\mathrm{clump,tot,obs}}$), the cloud mass ($M_{\mathrm{cloud}}$) and the SFR of the cloud ($\mathrm{SFR}_{\mathrm{cloud,obs}}$) in the inner Milky Way. The red dashed line represents a linear fit to the scatter points, with $k$ and $r$ denoting the slope and the correlation coefficient. The contours show the density distribution of the scatter points, derived from a Kernel Density Estimate.}
\label{mw}
\end{figure}

In \citet{bridge}, we identified molecular clouds in the inner Milky Way using CO (1–0) data cubes and cross-matched them with the ATLASGAL clumps cataloged by \citet{Urquhart2022-510}. 
Then, we calculated the star formation rate surface density of each cloud using the {\it Spitzer} 24 $\mu$m image,
\begin{equation}
    \frac{\Sigma_{\rm SFR}}{\rm M_{\odot}~yr^{-1}~kpc^{-2}} = 3.8\times10^{-3} \left(\frac{I_{\rm 24 \mu m}}{\rm MJy~sr^{-1}}\right),
\end{equation}
\begin{equation}
    \mathrm{SFR} = \Sigma_{\rm SFR} \times A
    \label{21um}
\end{equation}
where $I_{\rm 24 \mu m}$ is the total 24 $\mu$m flux contained within the cloud of area $A$. 
Finally, we found strong correlations between the total clump mass in a cloud ($M_{\mathrm{clump,tot,obs}}$), the cloud mass ($M_{\mathrm{cloud}}$) and the SFR of the cloud ($\mathrm{SFR}_{\mathrm{cloud,obs}}$), as shown in Fig.\ref{mw}. The correlations are
\begin{equation}
\begin{split}
&\mathrm{log}_{\rm 10} (M_{\mathrm{clump,tot,obs}}/M_{\odot}) = \\ 
&(0.75 \pm 0.02) \times \mathrm{log}_{\rm 10} (M_{\mathrm{cloud}}/M_{\odot}) - (0.56 \pm 0.13),
\end{split}
\label{Mcto}
\end{equation}

\begin{equation}
\begin{split}
&\mathrm{log}_{\rm 10} (\mathrm{SFR}_{\mathrm{cloud,obs}}/(M_{\odot}~{\rm yr}^{-1})) = \\ &(0.98 \pm 0.01) \times \mathrm{log}_{\rm 10} (M_{\mathrm{cloud}}/M_{\odot}) - (9.12 \pm 0.06).
\end{split}
\label{sfr-cloud}
\end{equation}

In \citet{Zhou2024PASP-2}, for the ATLASGAL clumps, we found that the clump star formation efficiency ($\mathrm{SFE}_{\mathrm{clump}}$) decreases with increasing clump mass ($M_{\mathrm{clump}}$), with a median value of $\approx$0.3. 
We also derived the correlations between the $\mathrm{SFE_{\mathrm{clump}}}$, the clump mass and the embedded cluster mass ($M_{\mathrm{ecl}}$), 

\begin{equation}
\begin{split}
&\mathrm{log}_{10} (\mathrm{SFE_{\mathrm{clump}}}) = \\ &(-0.37 \pm 0.01) \times \mathrm{log}_{10} (M_{\mathrm{clump}}/M_{\odot}) + (0.42 \pm 0.04),
\label{sfe-mcl}
\end{split}
\end{equation}
\begin{equation}
\begin{split}
&\mathrm{log}_{10} (M_{\mathrm{clump}}/M_{\odot}) = \\ &(1.02 \pm 0.02) \times \mathrm{log}_{10} (M_{\mathrm{ecl}}/M_{\odot}) + (0.52 \pm 0.05).
\label{mecl}
\end{split}
\end{equation}

\subsection{Clump mass function}\label{CLMF}

Given that clumps represent the progenitors of embedded star clusters, we assume a direct inheritance relationship between the clump mass function (CLMF) and the embedded cluster mass function (ECMF).
Then we adopt the formalism summarized in \citet{Yan2017-607}.
The CLMF is a single slope power law with a variable power-law index, $\beta$,
\begin{equation}\label{eq:xi_ecl}
\xi_{\mathrm{clump}}(M)=
k_{\mathrm{clump}} \times M^{-\beta}, M_{\mathrm{min}} \leqslant M<M_{\mathrm{max}},
\end{equation}
where $M_{\mathrm{min}}$ is the assumed lower limit of clump masses, $M_{\mathrm{max}}$ is the upper integration limit in the optimal sampling method defined in \citet{Schulz2015-582} , and $k_{\mathrm{clump}}$ is a normalization constant. The mass of the smallest stellar group observed in \cite{Kuhn2015-812,Zhou2024-688L} is $\approx 5 M_{\odot}$. 
When the uncertainties in both equations \ref{sfe-mcl} and \ref{mecl} are taken into account, the resulting SFE lies in the range [0.83, 1]. In the limiting case of SFE=1, the minimum clump mass is also 5 $M_{\odot}$. In order to guarantee sufficient sampling, we set the minimum clump mass to $M_{\mathrm{min}}= 5 M_{\odot}$.
Generally, the slope of the star cluster mass function is $\approx$ -2 \citep{Lada2003-41,Krumholz2019-57,Mok2020-893,Wainer2022-928,Zhou2024-688L,Zhou2025-541}. In \citet{Elia2017-471}, for Hi-GAL (Herschel InfraRed Galactic Plane Survey) clumps in the inner Milky Way, the slopes of the clump mass functions are between -1.88 and -2.46.

The parameters $k_{\mathrm{clump}}$ and $M_{\mathrm{max}}$ in equation \ref{eq:xi_ecl} are determined by solving equations \ref{Mctt} and \ref{eq:1intMecl} together, i.e., by invoking the clump population mass conservation:
\begin{equation}\label{Mctt}
M_{\mathrm{clump,tot}}=\int_{M_{\mathrm{min}}}^{M_{\mathrm{max}}}M~\xi_{\mathrm{clump}}(M)\,\mathrm{d}M.
\end{equation}
The optimal sampling normalization condition:
\begin{equation}\label{eq:1intMecl}
1=\int_{M_{\mathrm{max}}}^{10^9~\mathrm{M}_{\odot}}\xi_{\mathrm{clump}}(M)\,\mathrm{d}M,
\end{equation}
see \citet{Yan2017-607} for details on the adopted upper integration limit of $10^9$ $M_{\odot}$. Actually, as long as the upper integration limit is much larger than the real $M_{\mathrm{max}}$, the result will not be affected. In Section \ref{s1}, the mass of the most massive clump is only $10^{4.6}$ $M_{\odot}$. 

\subsection{Initial total mass of clumps}\label{s-initial}

Clumps, which are self-gravitating structures within molecular clouds \citep{Liu2016-829, Urquhart2018-473, Evans2021-920}, evolve more rapidly than the clouds themselves. By the time large-scale molecular clouds condense into dense, localized regions within galaxies, star formation has often already taken place within these regions. The molecular clouds we observe are typically composites, containing both
exposed and embedded stellar populations, as well as clumps \citep{Kuhn2020-899,Turner2022-516,Zhou2025-541}. Nearly all of those stellar populations—whether visible or still embedded—originate from clumps. These clumps may either be currently existing or have existed in the past. Past clumps are now observed as exposed and embedded stellar populations. 
Together, existing and past clumps constitute the initial clump population of the cloud, described by the initial clump mass function introduced in Section \ref{CLMF}. 
Therefore, $M_{\mathrm{clump,tot}}$ in equation \ref{Mctt} and $M_{\mathrm{clump,tot,obs}}$ in equation \ref{Mcto} are different. $M_{\mathrm{clump,tot}}$ includes both past and present clumps, while $M_{\mathrm{clump,tot,obs}}$ includes only the clumps currently observed.
As an estimate, 
\begin{align}
M_{\mathrm{clump,tot}} 
&= M_{\mathrm{clump,tot,obs}} +
   (M_{\mathrm{emb,tot,obs}} + M_{\mathrm{exp,tot,obs}})/0.3 \nonumber \\
&= M_{\mathrm{clump,tot,obs}} + 
   M_{\mathrm{cloud}} \times \mathrm{SFE}_{\mathrm{cloud}} / 0.3,
\label{Mcti}
\end{align}
where $M_{\mathrm{emb,tot,obs}}$ and $M_{\mathrm{exp,tot,obs}}$ are the total masses of exposed and embedded stellar populations in the cloud, respectively. The median value of the clump-scale SFE is 0.3 \citep{Zhou2024PASP-2}.
We note that the clump-scale SFE decreases with increasing clump mass; therefore, adopting a median value of 0.3 would lead to an overestimate of the total clump mass.
The SFE of the cloud ($\mathrm{SFE}_{\mathrm{cloud}}$) is defined as, ($M_{\mathrm{emb,tot,obs}}$+$M_{\mathrm{exp,tot,obs}}$)/$M_{\mathrm{cloud}}$, 
as calculated in \citet{Zhou2025-541}, 
\begin{equation}
\begin{split}
&\mathrm{log}_{\rm 10} (\mathrm{SFE}_{\mathrm{cloud}}) = \\ &(-0.32 \pm 0.06) \times \mathrm{log}_{\rm 10} (M_{\mathrm{cloud}}/M_{\odot}) - (0.034 \pm 0.37).
\end{split}
\label{sfe-cloud}
\end{equation}

\subsection{Star formation history}\label{SFH}

Equation \ref{eq:xi_ecl} defines the initial clump mass function of the cloud. 
The theoretically initial clump population has now evolved into a mixture of the currently observed clumps and the embedded/exposed stellar populations. The currently observed clumps correspond to the youngest ones, while the embedded and exposed stellar populations can be regarded as the intermediate-age and oldest clumps, respectively.  
The different components in a molecular cloud — clumps, and embedded and exposed stellar populations — can be regarded as representing the age spread of clumps within the cloud. 
We then model the star formation history (SFH) of the cloud to distinguish between these different components.
Using the SFH models presented in \citet{Zhou2024PASP-1,Dib2025-693},
we examine scenarios where the SFH remains constant and others where it varies over time. Through the SFH,
we create an age distribution of the initial clump population in a cloud to assign an age to each clump. Then, we sort out the intermediate-age clumps, which represent the embedded stellar populations and mainly contribute to the mid-infrared emission of the cloud.

The typical survival timescale of GMCs ($t_{\rm GMC}$) is 10$-$30 Myr \citep{Chevance2021,Kim2021-504,Kim2022-516}. 
In the cases of a constant SFH, we randomly sample the age of clumps, $t_{\rm b}$, with a uniform probability in the age range [0, $t_{\rm GMC}$] Myr. 
For a time-varying SFH, we employ a Gaussian function, allowing us to adjust both the peak position and the distribution width in time. The clump age, $t_{\rm b}$, is then randomly drawn from the corresponding Gaussian distribution,
\begin{equation}
P(t_{\rm b}) = \frac{1}{\sigma_{\rm t} \sqrt{2 \pi}} \exp\left[-\frac{1}{2} \left(\frac{t_{\rm b}-t_{\rm b,p}}{\sigma_{\rm t}}\right)^2 \right],
\label{eq4}
\end{equation}
where $t_{\rm b,p}$ is the peak position, and $\sigma_{\rm t}$ is the standard deviation.
We should consider different values of $\sigma_{\rm t}$ to cover possible time spans of star formation.
We consider values of $t_{\rm b,p}= 2, 4, 6, 8$ Myr and $\sigma_{\rm t}=$ 3, 6 and 9 Myr. There are a total of 12 cases here. Actually, the time-dependent SFH with large $\sigma_{\rm t}$ is similar to the constant SFH.
From a clump to an embedded cluster, the formation time is $\approx$2 Myr \citep{Evans2009-181,Covey2010-722,Megeath2022-134,Wells2022-516,Kim2023-944}.
Thereafter, the duration time of the embedded phase ($t_{\rm emb}$) is $\approx$2-7 Myr \citep{Kim2021-504,Kim2023-944}. 
Therefore, we only 
select the clumps in the age range [2, $t_{\rm emb}$+2] Myr to calculate the SFR and compare with the observation. Equation \ref{mecl} is used to convert the clump mass to the embedded cluster mass. Then the total embedded cluster mass is divided by $t_{\rm emb}$ to predict the SFR of the cloud ($\mathrm{SFR}_{\mathrm{cloud,p}}$).

\section{Results and discussion}\label{result}

\subsection{Procedure}\label{procedure}

The predictions of the model account for the cumulative effect of uncertainties across the entire modeling chain. 
Rather than providing single-point estimates, the model generates probability distributions for SFR predictions, acknowledging that our knowledge of physical parameters is inherently uncertain. 
The model adopts a nested Monte Carlo framework in which molecular cloud masses are sampled over a logarithmic grid from $10^3$ to $10^8$ M$_\odot$ (1000 points). For each cloud mass, multiple Monte Carlo realizations (typically 100) are performed, with all model parameters independently resampled in each iteration. This design enables a systematic assessment of how uncertainties propagate across the cloud mass spectrum. The resulting ensemble of SFR predictions for each mass is used to derive statistical descriptors, including the mean and median as central estimates and the standard deviation as a measure of dispersion. A key strength of this approach is the full propagation of uncertainties through all stages of the calculation, naturally accounting for non-linear effects and parameter correlations that are not captured by simplified analytic error-propagation methods.

The logic and procedures of the model are as follows:
(1) Use equation \ref{Mcti} to estimate the initial total mass of clumps in a cloud;
(2) Generate the initial clump population of the cloud using the initial clump mass function;
(3) Model the SFH of the cloud to assign an age to each clump;
(4) Identify the clumps in the age range [2, $t_{\rm emb}$+2] Myr and calculate the total embedded cluster mass;
(5) Predict the SFR based on the duration of the embedded phase.
Some details should be emphasized here:
(1) We sample $\beta$ uniformly between 1.88 and 2.46 in each Monte Carlo iteration. This reflects the observed variation in CLMF slopes across different molecular clouds and star-forming regions. The uniform distribution represents a conservative assumption of maximum uncertainty within the observed range;
(2) The maximum age for star formation history is sampled uniformly between 10 and 30 Myr, reflecting the typical survival timescale of GMCs. The starting point of the embedded phase is fixed at 2 Myr, representing the time required for a clump to evolve into an observable embedded cluster. The endpoint of the embedded phase is sampled uniformly between 4 and 9 Myr, since the duration time of the embedded phase is $\approx$2-7 Myr; 
(3) All empirical relations derived from observations in Section \ref{model} are parameterized with truncated normal priors on their slopes and intercepts, capturing empirical uncertainties while enforcing physically meaningful bounds.

The model incorporates multiple internal consistency checks. Mass conservation verifies that the generated clump masses sum to the specified total mass within tolerance. Physical plausibility checks ensure that converted embedded cluster masses never exceed their parent clump masses. Age distribution checks confirm that assigned ages respect the specified SFH model.
For each molecular cloud mass, we monitor convergence of the Monte Carlo statistics as iterations progress. We check stabilization of the mean, standard deviation, and percentile estimates to ensure sufficient sampling of the parameter space. 
The model reports the fraction of successful iterations (those completing without numerical errors) as a quality metric, see the online implementation codes for more details \footnote{\url{https://github.com/jianwenzhou11/SFR_model.git}}.

\subsection{Transition mass}\label{prediction}

\begin{figure}
\centering
\includegraphics[width=0.475\textwidth]{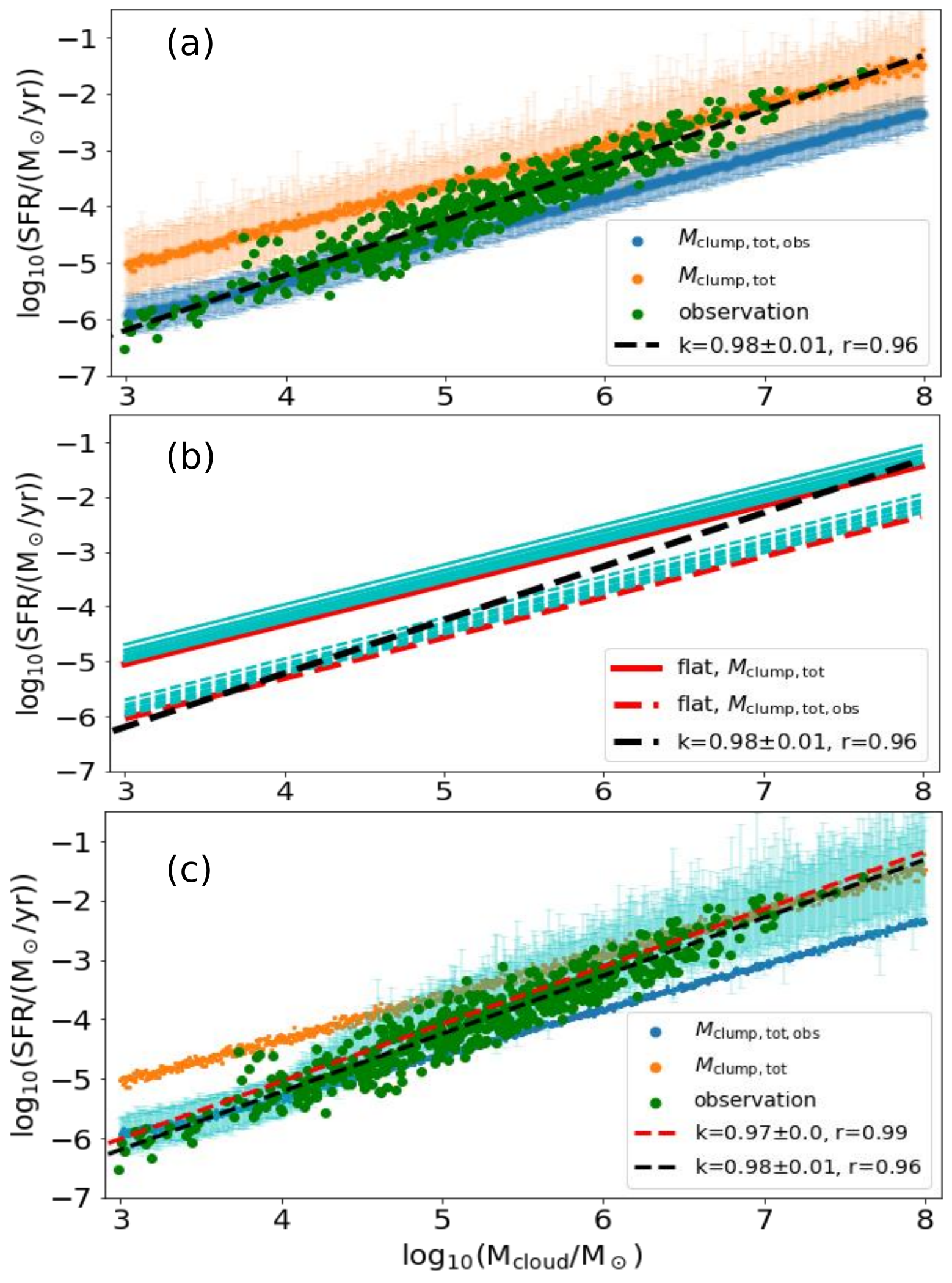}
\caption{
Compare the observed SFR with the model-predicted SFR. Green points represent the observation. The black dashed line shows the linear fit to the green points, with $k$ and $r$ denoting the slope and the correlation coefficient, respectively. (a) The orange and blue points are the predictions under the constant SFH  based on equation \ref{Mcti} and equation \ref{Mcto}, respectively, and the shaded regions indicate the uncertainty ranges; (b) Solid and dashed lines indicate the linear fits of the predictions for the 13 different SFHs, with the red line highlighting the result for the constant SFH. The black dashed line represents the observation; (c) Same as panel (a). The red dashed line shows a fit to the predictions based on equation \ref{Mctio}, while the cyan shaded region indicates the uncertainty in the predicted results.}
\label{pred}
\end{figure}

Fig.\ref{pred}(a) shows the results under a constant SFH. At the high-mass end, $\mathrm{SFR}_{\mathrm{cloud,p}}$ and $\mathrm{SFR}_{\mathrm{cloud,obs}}$ are comparable, but $\mathrm{SFR}_{\mathrm{cloud,p}}$ is significantly larger than $\mathrm{SFR}_{\mathrm{cloud,obs}}$ at the low-mass end. If we change equation \ref{Mcti} into equation \ref{Mcto}, the result is opposite. All 13 SFHs (12 Gaussian and 1 constant) are compared in Fig.\ref{pred}(b), and the constant SFH simultaneously provides the best fit to both the low-mass and high-mass ends. Therefore, we focus exclusively on the constant SFH in the following discussion.

The discussion in Section \ref{s-initial} mainly concerns molecular clouds at the high-mass end, which can survive for a long time despite hosting stellar populations. In contrast, in low-mass molecular clouds, feedback from the stellar populations can efficiently disrupt the cloud. 
Hence, observed low-mass molecular clouds are not expected to contain a significant fraction of stellar components. Actually, equation \ref{sfe-cloud} from \citet{Zhou2025-541} is only for clouds with masses larger than 
$10^{5.5}$ $M_{\odot}$. 
Therefore, we need a new relation: it should follow equation \ref{Mcti} for high-mass molecular clouds, equation \ref{Mcto} for low-mass molecular clouds, and lie between the predictions of the two equations for intermediate-mass molecular clouds.
To obtain a smooth transition between the two regimes, we adopt a linear weighting function defined as
\begin{equation}
t = \frac{\mathrm{log}_{\rm 10} (M_{\mathrm{cloud}}/M_{\odot}) - a}{b - a} , 
\end{equation}
where $a$ and $b$ specify the transition mass range. 
The parameter $t$ is limited to the range [0,1].
The final relation is then given by
\begin{equation}
M_{\mathrm{clump,tot,m}} = (1 - t)\, M_{\mathrm{clump,tot,obs}} + t\, M_{\mathrm{clump,tot}}~,
\label{Mctio}
\end{equation}
where $M_{\mathrm{clump,tot,obs}}$ and $M_{\mathrm{clump,tot}}$ correspond to the relations of 
equation \ref{Mcto} and equation \ref{Mcti}, respectively. 
Using only equation \ref{Mcti} or equation \ref{Mcto} represents two extreme cases. 
A grid of molecular cloud masses spanning the full range of interest ([$10^3$, $10^8$] M$_\odot$) is constructed, and for each mass, Monte Carlo simulations are performed assuming purely equation \ref{Mcto} or purely equation \ref{Mcti} contributions.
By fitting the two extremes, we estimate the critical masses at which the model predictions begin to significantly deviate from the observation and then perform local optimization starting from these initial values. 
The optimization seeks to drive the median of the ratio between the predicted and target SFRs toward unity while minimizing its variance. The loss function combines an absolute-deviation term with a standard-deviation penalty:
\begin{equation}
\mathcal{L} = |R_{\text{median}} - 1| +  \frac{\sigma_R}{R_{\text{median}}} , 
\end{equation}
where $R_{\text{median}}$ is the median SFR ratio across all Monte Carlo realizations, and $\sigma_R$ is the standard deviation of the SFR ratio.
Finally, we obtain $a \approx 4$ and $b \approx 6.5$. 
For molecular clouds in the Milky Way, in \cite{Lee2016-833}, the total stellar mass contained in these clouds is inferred from free–free fluxes, with the least massive clouds having a mass of about 10$^{4}$ $M_{\odot}$. This fact supports $a \approx$ 4.
The combination of equation.\ref{Mcto} and equation.\ref{Mcti} also partially compensates for the overestimation of the total clump mass caused by adopting a value of 0.3 for the clump-scale SFE. 
As shown in Fig.\ref{pred}(c), after the combination of equation \ref{Mcti} and equation \ref{Mcto}, now the model can well fit the observation.

\subsection{Uncertainty}

\begin{table}
\centering
\caption{Parameter-specific contributions to the total uncertainty ranked by $\overline{\text{CV}}_i$ in descending order. }
\label{pc}
\begin{tabular}{lcccc}
\hline
Parameter & $\overline{\text{CV}}_i$ & $\overline{\Delta}_i$ & $f_i$ \\
\hline
equation \ref{sfe-cloud} & 0.62 & 0.37 & 15.94 \\
equation \ref{sfe-cloud}\_slope     & 0.60 & 0.36 & 15.42 \\
$t_{\rm GMC}$        & 0.43 & 0.36 & 11.05 \\
equation \ref{Mcto}\_slope     & 0.30 & 0.32 & 7.71 \\
equation \ref{mecl}\_intercept & 0.30 & 0.28 & 7.71 \\
$t_{\rm emb}$      & 0.29 & 0.33 & 7.46 \\
equation \ref{mecl}\_slope     & 0.28 & 0.31 & 7.19 \\
equation \ref{sfr-cloud}\_intercept & 0.27 & 0.28 & 6.94 \\
$\beta$          & 0.27 & 0.28 & 6.94 \\
equation \ref{Mcto}\_intercept & 0.27 & 0.30 & 6.94 \\
equation \ref{sfr-cloud}\_slope     & 0.26 & 0.26 & 6.68 \\
\hline
\end{tabular}
\end{table}

The uncertainty analysis is based on a One-at-a-Time (OAT) methodology, with additional consideration of statistical sampling and parameter distributions. Before evaluating the contributions of individual parameters, a reference is defined using the median values of all parameters. Specifically, for the reference, we neglect uncertainties in all empirical relations, and for parameters defined within a finite range, we adopt the midpoint of the allowed interval. Each parameter $p_i$ is then examined independently, while all other parameters are held fixed at their reference values. This method can decompose the total uncertainty in SFR predictions into contributions from individual parameters and physical processes. It is crucial for identifying which sources of uncertainty dominate the model's predictive variance and guiding efforts to improve model precision.

For each parameter $p_i$ and each cloud mass $M_k$, the set of SFR predictions $\{\text{SFR}_{i}^{(j)}(M_k)\}_{j=1}^{N}$ is analyzed:

\begin{align*}
\mu_i(M_k) &= \frac{1}{N} \sum_{j=1}^{N} \text{SFR}_{i}^{(j)}(M_k) ,\\
\sigma_i(M_k) &= \sqrt{\frac{1}{N-1} \sum_{j=1}^{N} \left(\text{SFR}_{i}^{(j)}(M_k) - \mu_i(M_k)\right)^2} ,\\
\text{CV}_i(M_k) &= \frac{\sigma_i(M_k)}{\mu_i(M_k)} \quad \text{(for $\mu_i > 0$)} ,\\
\Delta_i(M_k) &= \frac{|\mu_i(M_k) - \text{SFR}_{\text{ref}}(M_k)|}{\text{SFR}_{\text{ref}}(M_k)} ,
\end{align*}
\noindent 
where $i$ denotes the $i$-th parameter in the analysis, and $j$ indexes the $j$-th random sample drawn for parameter $i$ at cloud mass $M_k$. The quantity $\mathrm{SFR}_{i}^{(j)}(M_k)$ denotes the predicted star formation rate obtained when parameter $i$ is set to its $j$-th randomly sampled value, while all other parameters are held fixed at their reference values. The symbol $N$ represents the Monte Carlo sample size.
Then the statistics are integrated across the cloud mass range:

\begin{equation}
\overline{\text{CV}}_i = \frac{1}{N_{\text{Mcloud}}} \sum_{k=1}^{N_{\text{Mcloud}}} \text{CV}_i(M_k) ,
\end{equation}

\begin{equation}
\overline{\Delta}_i = \frac{1}{N_{\text{Mcloud}}} \sum_{k=1}^{N_{\text{Mcloud}}} \Delta_i(M_k) .
\end{equation}

\noindent Parameter-specific contributions to the total uncertainty are then ranked by $\overline{\text{CV}}_i$ in descending order. The statistical measure
$\overline{\text{CV}}_i$ is prioritized over $\overline{\Delta}_i$ because it directly quantifies the prediction uncertainty range—how much SFR predictions scatter due to parameter variations—whereas $\overline{\Delta}_i$ only measures systematic bias from a reference. In uncertainty analysis, understanding the full spread of possible outcomes (captured by $\overline{\text{CV}}_i$) is more critical than knowing the average offset, as symmetric distributions can show zero $\overline{\Delta}_i$ while still having substantial uncertainty. The scale-invariance of $\overline{\text{CV}}_i$ also allows consistent comparison across parameters despite the five-order magnitude range in $M_{\rm cloud}$, making it the appropriate metric for ranking parameter contributions to overall model uncertainty.

The final step decomposes total uncertainty into contributions from individual parameters, employing the coefficient of variation as a dimensionless and scale-invariant measure:
\begin{equation}
f_i = \frac{\overline{\text{CV}}_i}{\sum_{j=1}^{N_{\text{p}}} \overline{\text{CV}}_j},
\end{equation}
where $N_{\text{p}}$ is the number of parameters.
As reflected in Table.\ref{pc}, equation \ref{sfe-cloud} contributes $\approx$31 \% of the total uncertainty, due to the large scatter in the molecular cloud SFE as a function of cloud mass \citep{Zhou2025-541}. 
The significant uncertainty introduced by equation \ref{sfe-cloud} is also clearly reflected in Fig.\ref{pred}(a), where the orange-shaded uncertainty band is substantially wider than the blue-shaded band.

\subsection{Current Limitations and Caveats}

In the model,
we considered three components within the molecular cloud, namely clumps, embedded and exposed stellar populations. The exposed stellar populations have already separated from the gas and therefore do not significantly affect the physical properties of the clumps. As discussed in \citet{Watkins2019-628A,Zhou2024-682-128,Zhou2024-682-173}, although the embedded stellar populations are spatially closely connected to the clumps, the feedback driven by the embedded stellar populations do not significantly influence the clumps’ physical properties. 
The reason lies in the hierarchical structure of the molecular gas, as described in Section \ref{sec:intro}.
The network structure of molecular clouds \citep{Zhou2025-699} implies that the knots or local dense structures as local hubs or local gravitational centers are relatively independent of each other. Although some knots in the cloud evolve more rapidly, forming embedded stellar populations, due to their relative independence from the neighboring
knots, early feedback from them does not significantly impact the physical properties of the neighboring knots (clumps). 


The uniform distributions for some parameters ($\beta$, $t_{\rm GMC}$, and $t_{\rm emb}$) represent maximum uncertainty assumptions rather than informed priors. 
The single-slope power-law CLMF may oversimplify the true mass distribution of clumps. The fixed starting point of the embedded phase at 2 Myr neglects possible environmental variations. 
In Section \ref{prediction}, the total clump mass predicted from the cloud mass significantly affects the results. To accurately estimate the initial total clump mass within a molecular cloud, we need precise measurements of both the current total clump mass and the total stellar mass contained within the cloud, especially for low-mass clouds. Equation \ref{Mctio} currently represents a compromise, and the exact relationship still needs to be constrained by observations. 

Anyway,
this model primarily provides a theoretical framework that integrates a wide range of observational results, thereby clarifying the tasks for future observations. At a minimum, observations are needed to constrain a set of model parameters that exhibit significant uncertainties, as shown in Table.\ref{pc}.

\section{Conclusion}\label{conclusion}

In the model, we compile empirical relations at the molecular cloud and clump scales, derived from observations of the Milky Way and nearby galaxies. Based on these relations, we construct the initial clump mass function (CLMF) of molecular clouds and model their star formation histories (SFHs).
We then employ a nested Monte Carlo framework to predict cloud-scale star formation rates, fully propagating uncertainties from all model parameters.
Cloud masses are sampled logarithmically from $10^3$ to $10^8~M_\odot$, and for each mass, multiple realizations resample parameters independently, producing probability distributions of SFRs rather than single-point estimates. Statistical descriptors, including the mean, median, and standard deviation, are used to quantify the central predictions and associated uncertainties.
Parameters such as the CLMF slope ($\beta$), GMC lifetime ($t_{\rm GMC}$), and embedded phase duration ($t_{\rm emb}$) are sampled from uniform distributions to account for observed variations, while empirical relations are assigned truncated normal priors to capture measurement uncertainties and enforce physically meaningful bounds.

Comparison with observations indicates that a constant SFH provides the best overall fit at both low-mass and high-mass ends. A smooth transition function (equation \ref{Mctio}) interpolates between these regimes by combining predictions from equations \ref{Mcto} and \ref{Mcti} (two extreme cases), with local optimization identifying the transition mass range, $\log_{10} (M_{\rm cloud}/M_\odot) \sim 4$–6.5.
Uncertainty analysis, based on a One-at-a-Time (OAT) methodology, decomposes the total SFR variance into parameter-specific contributions using the coefficient of variation ($\overline{\rm CV}_i$). Table.\ref{pc} shows that the cloud-scale star formation efficiency (equation \ref{sfe-cloud}) accounts for $\approx$ 31\% of the total uncertainty, due to its weak correlation with cloud mass.

Current limitations of the model include simplifying assumptions regarding clump independence, the adoption of a single-slope CLMF, uniform distributions for key parameters, and a fixed starting time for the embedded phase. Observational constraints on total clump and stellar masses—particularly in low-mass clouds—remain critical for refining predictions. Nevertheless, the model provides a coherent framework linking cloud and clump properties to cloud-scale SFRs and helps guide future observations toward parameters that significantly contribute  to predictive uncertainty.

\section*{Acknowledgements}
Thanks to the referee for the detailed and constructive comments, which have significantly contributed to improving this work. Thanks to Zhiqiang Yan for the helpful discussion.

\section{Data availability}
All data used in this work are available from the first author upon request. The implementation code and its documentation are available online
\footnote{\url{https://github.com/jianwenzhou11/SFR_model.git}}.

\bibliography{ref}
\bibliographystyle{aasjournal}

\begin{appendix}



\end{appendix}

\end{document}